\documentclass[eqsecnum,preprint,prd,aps,nofootinbib]{revtex4}
\usepackage{amsmath}
\usepackage{graphics}
\def\lvec#1{\vbox{\ialign{##\crcr$\leftarrow$\crcr\noalign{
 \kern-1pt\nointerlineskip}$\hfil\displaystyle{#1}\hfil$\crcr}}}
\def\rvec#1{\vbox{\ialign{##\crcr$\rightarrow$\crcr\noalign{
 \kern-1pt\nointerlineskip}$\hfil\displaystyle{#1}\hfil$\crcr}}}
\begin{document}
\begin{center}
\bibliographystyle{article}

{\Large \textsc{A new antisymmetric bilinear map for type-I gauge theories}}

\end{center}
\vspace{0.4cm}


\date{\today}

\author{Giampiero Esposito,$^{1,2}$ \thanks{%
Electronic address: giampiero.esposito@na.infn.it} 
Cosimo Stornaiolo$^{1,2}$ \thanks{%
Electronic address: cosmo@na.infn.it}}
\affiliation{${\ }^{1}$Istituto Nazionale di Fisica Nucleare, 
Sezione di Napoli,\\
Complesso Universitario di Monte S. Angelo, Via Cintia, Edificio 6, 80126
Napoli, Italy\\
${\ }^{2}$Dipartimento di Scienze Fisiche, Complesso Universitario di Monte
S. Angelo,\\
Via Cintia, Edificio 6, 80126 Napoli, Italy}

\begin{abstract}
In the case of gauge theories, which are ruled by an infinite-dimensional
invariance group, various choices of antisymmetric bilinear maps on field
functionals are indeed available. This paper proves first that, within
this broad framework, the Peierls map (not yet the bracket) is a member
of a larger family. At that stage, restriction to gauge-invariant
functionals of the fields, with the associated Ward identities and geometric
structure of the space of histories, make it possible to prove that the
new map is indeed a Poisson bracket in the simple but relevant case of
Maxwell theory. The building blocks
are available for gauge theories only: vector fields that leave the action
functional invariant; the invertible gauge-field operator, and
the Green function of the ghost operator.
\end{abstract}
\maketitle
\bigskip
\vspace{2cm}

\section{Introduction}

In the global approach to quantum field theory \cite{DeWi03}, the physical
arena is given by the infinite-dimensional manifold $\Phi$ of field
histories $\varphi^{i}$. For gauge fields, there exists on $\Phi$ a set of
vector fields $Q_{\alpha}$ that leave the action $S$ invariant, i.e.
\begin{equation}
Q_{\alpha}S=0.
\label{(1.1)}
\end{equation}
On denoting by $Q_{\; \alpha}^{i}$ the components of the $Q_{\alpha}$, one
can write Eq. (1.1) in the form
\begin{equation}
S_{,i}Q_{\; \alpha}^{i}=0,
\label{(1.2)}
\end{equation}
the comma being hereafter the standard notation for functional derivatives
${\delta \over \delta \varphi^{i}}$ with respect to field 
variables \cite{DeWi03}.
The vector fields $Q_{\alpha}$ are linearly independent, and for type-I
gauge theories the Lie brackets of the $Q_{\alpha}$ depend linearly on
the $Q_{\alpha}$ themselves:
\begin{equation}
\Bigr[Q_{\alpha},Q_{\beta}\Bigr]=C_{\; \alpha \beta}^{\gamma} \; Q_{\gamma},
\label{(1.3)}
\end{equation}
where the $C_{\; \alpha \beta}^{\gamma}$ are independent of field 
variables: $C_{\; \alpha \beta , i}^{\gamma}=0$, and are therefore
called `structure constants'. The {\it proper gauge group} ${\cal G}$ is 
the set of transformations of $\Phi$ into itself obtained by exponentiating
the infinitesimal gauge transformation
\begin{equation}
\delta \varphi^{i}=Q_{\; \alpha}^{i} \; \delta \xi^{\alpha},
\label{(1.4)}
\end{equation}
and taking products of the resulting exponential maps \cite{DeWi03}. 
The closure property expressed by Eq. (1.3) implies that the proper gauge
group decomposes $\Phi$ into subspaces, known as orbits, to which the
$Q_{\alpha}$ are tangent. The space of orbits is the quotient 
space $\Phi / {\cal G}$. 

In recent work by the authors \cite{Espo07}, 
we have considered the problem of defining the most general Poisson 
bracket on the space of gauge-invariant functionals for type-I theories.
Our original hope was that, in the same way as the Moyal bracket
\cite{Moya49, Jord61, Grac02} is proportional to the Poisson bracket to lowest
order in $\hbar$, one might be able to define a new Poisson bracket
that is proportional to the Peierls bracket 
\cite{Peie52} to lowest order in $\hbar$. 
In the latter, the building block is the {\it supercommutator function}, 
i.e. the difference 
\begin{equation}
{\widetilde G}^{jk} \equiv G^{+jk}-G^{-jk}=-{\widetilde G}^{kj}
\label{(1.5)}
\end{equation}
of advanced and retarded Green functions of the invertible operator 
$F_{ij}$ on gauge fields $\varphi^{j}$ obtained after adding a
gauge-breaking term in the functional integral \cite{DeWi03}. 
[By virtue of the definition (1.5), the integral representation of
${\widetilde G}^{jk}$ is given by a contour in the complex
$p^{0}$-plane passing below the poles on the real line] 
We were thus looking for a quantum commutator having the following
asymptotic expansion:
\begin{equation}
[A,B] \sim i {\hbar}(A,B)+{\rm O}({\hbar}^{3}),
\label{(1.6)}
\end{equation}
where $A$ and $B$ are any two gauge-invariant functionals:
\begin{equation}
Q_{\alpha}A=Q_{\alpha}B=0,
\label{(1.7)}
\end{equation}
and $(A,B)$ is their Peierls bracket \cite{DeWi03, Espo07}
\begin{equation}
(A,B) \equiv A_{,i}{\widetilde G}^{ij}B_{,j}
=\int \int dx \; dy {\delta A \over \delta \varphi^{i}(x)}
{\widetilde G}^{ij}(x,y){\delta B \over \delta \varphi^{j}(y)}.
\label{(1.8)}
\end{equation}
With hindsight, we were misled in putting the emphasis on Moyal
brackets, since the phase-space structure is totally extraneous
to a space-of-histories formulation. With this improved understanding,
section 2 builds a map, not yet a Poisson bracket, which is antisymmetric
and bilinear. Sections 3 and 4 
exploit the geometry of the space of histories
to give a concrete form to our new map, and show under which conditions
it becomes a Poisson bracket. Section 5 derives its gauge transformation
properties. Concluding remarks and open problems are
presented in section 6.

\section{An antisymmetric bilinear map}

Given the space of histories $\Phi$ for type-I theories, with the
associated space of functionals ${\cal F}(\Phi)$,
suppose we want to define a suitable map (not just its asymptotic
expansion)
$$
[\; , \;]:{\cal F}(\Phi) \times {\cal F}(\Phi) \rightarrow
{\cal F}(\Phi)
$$
acting on any two field functionals $A$ and $B$ according to
\begin{equation}
[A,B] \equiv i{\hbar} (A,B)+\mu_{3}(i{\hbar})^{3}
A_{,jkl}U^{jp}U^{kq}U^{lr}B_{,pqr}.
\label{(2.1)}
\end{equation}
Note that, at this stage, (2.1) is not a Poisson bracket, since
the Peierls map $(A,B)$ (whose action on $A$ and $B$ is analogous 
to (1.8)) is not a Poisson bracket for generic $A$ and $B$.
The coefficient $\mu_{3}$ of $(i{\hbar})^{3}$ should be so chosen that 
the two terms on the right-hand side of (2.1) have the same dimension.
However, we stress that the form of $U^{jp}$ is not fixed a priori. 
From the definition (2.1) we find
\begin{equation}
[B,A]=-i{\hbar}(A,B)+\mu_{3}(i{\hbar})^{3} B_{,pqr}
U^{pj}U^{qk}U^{rl}A_{,jkl},
\label{(2.2)}
\end{equation}
and hence antisymmetry of our map:
\begin{equation}
[A,B]=-[B,A],
\label{(2.3)}
\end{equation}
is achieved if and only if 
\begin{equation}
U^{jp}=-U^{pj},
\label{(2.4)}
\end{equation}
with $U^{jp}$ otherwise arbitrary for the time being. Moreover, 
bilinearity follows at once from the definition (2.1), i.e.
\begin{equation}
[A,B+C]=[A,B]+[A,C].
\label{(2.5)}
\end{equation}

\section{Geometry of the space of histories}

It is now helpful to summarize some key features of 
the geometry of the space $\Phi$ of histories.
For this purpose, let us recall that $\Phi$ is endowed with a 
gauge-invariant and ultra-local metric 
$\gamma$ \cite{DeWi03}. Gauge invariance 
means the vanishing of the Lie derivative of $\gamma$ along the vector 
fields $Q_{\alpha}$ that leave the action invariant as in Eq. (1.1), i.e.
\begin{equation}
L_{Q_{\alpha}}\gamma=0,
\label{(3.1)}
\end{equation}
while ultra-locality of $\gamma$ is expressed by
\begin{equation}
\gamma_{ij'}=\delta(x,x')f_{ij},
\label{(3.2)}
\end{equation}
with $f_{ij}$ independent of space-time derivatives of the fields
$\varphi^{i}$. One can therefore build the operator
\begin{equation}
{\cal F}_{\alpha \beta}=-Q_{\; \alpha}^{i} \; \gamma_{ij} \;
Q_{\; \beta}^{j},
\label{(3.3)}
\end{equation}
whose Green functions satisfy
\begin{equation}
{\cal F}_{\alpha \beta}G^{\beta \gamma}=-\delta_{\alpha}^{\; \gamma}.
\label{(3.4)}
\end{equation}
It is then possible to introduce a family of connection $l$-forms on
$\Phi$, i.e.
\begin{equation}
\omega_{\; i}^{\alpha} \equiv \gamma_{ij} \; Q_{\; \beta}^{j} \;
G^{\beta \alpha},
\label{(3.5)}
\end{equation}
which take values in the Lie algebra of the proper gauge group and 
play the role of `inverses' of the $Q_{\; \alpha}^{i}$ with
respect to field indices, in that
\begin{equation}
Q_{\; \alpha}^{i} \; \omega_{i}^{\; \beta}
=Q_{\; \alpha}^{i} \; \gamma_{ij} \; Q_{\; \gamma}^{j} \; G^{\gamma \beta}
=-{\cal F}_{\alpha \gamma}G^{\gamma \beta}=\delta_{\alpha}^{\; \beta}.
\label{(3.6)}
\end{equation}

The fibre-adapted coordinates available can be denoted by $I^{A}$ and
$K^{\alpha}$, where the $I$'s label the fibres, i.e. the points in
$\Phi / {\cal G}$, and are gauge-invariant, i.e. \cite{DeWi03}
\begin{equation}
Q_{\alpha}I^{A}=0,
\label{(3.7)}
\end{equation}
while the $K$'s label the points within each fibre, and correspond
to the choice of gauge-fixing functional in physical language. One usually
singles out a base point $\varphi_{\star}$ in $\Phi$ and chooses the
$K$'s to be local functionals of the $\varphi$'s of such a form that
the `matrix' \cite{DeWi03}
\begin{equation}
{\widehat F}_{\; \beta}^{\alpha}
=Q_{\beta}K^{\alpha}=K_{\; ,i}^{\alpha} \; Q_{\; \beta}^{i}
\label{(3.8)}
\end{equation}
is actually a non-singular differential operator at and in a neighbourhood
of $\varphi_{\star}$. The operator (3.8) is 
the ghost operator \cite{Fadd67, DeWi67} of modern quantum
field theory, and its Green functions are denoted by 
${\widehat G}^{\beta \gamma}$. The latter solve the equation
\begin{equation}
{\widehat F}_{\alpha \beta}{\widehat G}^{\beta \gamma}
=-\delta_{\alpha}^{\; \gamma},
\label{(3.9)}
\end{equation}
and can obey one of the various boundary conditions defining hyperbolic
Green functions (e.g. advanced, or retarded, or Feynman). 

The invertible gauge-field operator $F_{ij}$ mentioned in Sec. 1 can be
written in the form (hereafter we assume that the loop expansion of the
$\langle {\rm out} | {\rm in} \rangle$ amplitudes is performed 
\cite{DeWi03})
\begin{equation}
F_{ij}=S_{,ij}+K_{\; ,i}^{\alpha} \; \Omega_{\alpha \beta} \;
K_{\; ,j}^{\beta},
\label{(3.10)}
\end{equation}
where $\Omega_{\alpha \beta}$ is a non-singular $\varphi$-dependent
local distribution having the gauge transformation law \cite{Espo07}
\begin{equation}
\delta \Omega_{\alpha \beta}=\Omega_{\alpha \beta ,i}
Q_{\; \gamma}^{i} \delta \xi^{\gamma}
=-\Bigr(\Omega_{\delta \beta}C_{\; \gamma \alpha}^{\delta}
+\Omega_{\alpha \delta} C_{\; \beta \gamma}^{\delta}\Bigr)
\delta \xi^{\gamma}.
\label{(3.11)}
\end{equation}

\section{The new form of $U^{jp}$}

We now try to build the antisymmetric $U^{jp}$ in (2.1) in such a
way that it involves only the concepts defined so far, it yields 
a Poisson bracket when the space ${\cal F}(\Phi)$ of section 2 is
restricted to the space of gauge-invariant functionals satisfying
Eq. (1.7) (this restriction being the one for which also the Peierls
map becomes a Poisson bracket), and has well-defined gauge
transformation properties (see Sec. 5). 
By virtue of these requirements, we are led to consider 
\begin{equation}
U^{jp} \equiv Q_{\; \alpha}^{[j} \; F_{\; l}^{p]} \; Q_{\; \beta}^{l}
\; {\widehat G}^{\alpha \beta},
\label{(4.1)}
\end{equation}
while bearing in mind, from (2.1), also the Ward identities obtained
by functional differentiation of (1.7), i.e., for $Z=A,B$,
\begin{equation}
Z_{,ij}Q_{\; \alpha}^{i}+Z_{,i}Q_{\; \alpha ,j}^{i}=0,
\label{(4.2)}
\end{equation}
\begin{equation}
Z_{,ijk}Q_{\; \alpha}^{i}+Z_{,ij}Q_{\; \alpha,k}^{i}
+Z_{,ik}Q_{\; \alpha,j}^{i}+Z_{,i}Q_{\; \alpha,jk}^{i}=0,
\label{(4.3)}
\end{equation}
where (4.2) and (4.3) are just two of the infinitely many Ward
identities available. In light of (2.1) and (4.1), (4.3), we find
\begin{eqnarray}
\; & \; & A_{,jkl}U^{jp}U^{ks}U^{lr}B_{,psr} 
=- {1\over 2}\Bigr(A_{,jk}Q_{\; \alpha,l}^{j}
+A_{,jl}Q_{\; \alpha,k}^{j}\Bigr) F_{\; l}^{p} \; Q_{\; \beta}^{l} \;
{\widehat G}^{\alpha \beta} B_{,psr}U^{ks}U^{lr} \nonumber \\
&+& {1\over 2}A_{,jkl} F_{\; l}^{j} \; Q_{\; \beta}^{l}
{\widehat G}^{\alpha \beta}
\Bigr(B_{,ps}Q_{\; \alpha,r}^{p}+B_{,pr}Q_{\; \alpha,s}^{p}\Bigr)
U^{ks}U^{lr}.
\label{(4.4)}
\end{eqnarray}
We have here exploited the linear dependence of $Q_{\; \alpha}^{i}$
on field variables for all type-I theories \cite{DeWi03}, i.e.
\begin{equation}
Q_{\; \alpha,jk}^{i}=0.
\label{(4.5)}
\end{equation}
Equation (4.4) can be reduced to an equation where only one functional
derivative of $A$ and two functional derivatives of $B$ occur (or the
other way around), by virtue of the Ward identities (4.2) and (4.3),
but it already displays a very important property: our definition (4.1)
can be used to obtain an `addition' to the Peierls map. The term
(4.4) does not vanish for generic type-I theories (e.g. Yang--Mills
or general relativity), but it vanishes for Maxwell theory, where
$Q_{\; \alpha}^{i}$ reduces to \cite{DeWi03}
\begin{equation}
Q_{\mu}(x,x')=-\delta_{,\mu}(x,x'),
\label{(4.6)}
\end{equation}
which implies that the $Q_{\; \alpha}^{i}$ are independent of field 
variables for Maxwell theory, i.e. $Q_{\; \alpha,j}^{i}=0$ in this case.

\section{Gauge transformation law of $U^{jp}$}

Under the infinitesimal gauge transformations (1.4), the generators
$Q_{\; \alpha}^{i}$, gauge-field operator $F_{ij}$ and ghost Green 
function ${\widehat G}^{\alpha \beta}$ transform according to
\cite{DeWi03,Espo07}
\begin{equation}
\delta Q_{\; \alpha}^{i}=Q_{\; \alpha,r}^{i} \;
Q_{\; \gamma}^{r} \delta \xi^{\gamma},
\label{(5.1)}
\end{equation}
\begin{equation}
\delta F_{ij}=-\Bigr(F_{kj}Q_{\; \gamma,i}^{k}
+F_{ik}Q_{\; \gamma,j}^{k}\Bigr) \delta \xi^{\gamma},
\label{(5.2)}
\end{equation}
\begin{equation}
\delta {\widehat G}^{\alpha \beta}=\Bigr(C_{\; \gamma \delta}^{\alpha}
\; {\widehat G}^{\delta \beta}-{\widehat G}_{\; \delta}^{\alpha} \;
C_{\; \gamma}^{\delta \; \; \beta}\Bigr)\delta \xi^{\gamma}.
\label{(5.3)}
\end{equation}
Thus, the antisymmetric $U^{jp}$ defined in (4.1) has a well defined 
gauge transformation law. By virtue of (5.1)--(5.3), such a law can
be eventually cast in the form
\begin{eqnarray}
\delta U^{jp}&=&{1\over 2} \biggr \{ \Bigr(Q_{\; \alpha,r}^{j} \;
Q_{\; \gamma}^{r} \; F_{\; l}^{p}-Q_{\; \alpha,r}^{p} \;
Q_{\; \gamma}^{r} \; F_{\; l}^{j}\Bigr) Q_{\; \beta}^{l} \;
{\widehat G}^{\alpha \beta} \nonumber \\
&+& \biggr[-Q_{\; \alpha}^{j}\Bigr(F_{kl} \; Q_{\; \gamma ,}^{k \; \; \; p}
+F_{\; k}^{p} \; Q_{\; \gamma,l}^{k}\Bigr)
+Q_{\; \alpha}^{p}\Bigr(F_{kl} \; Q_{\; \gamma ,}^{k \; \; \; j}
+F_{\; k}^{j} \; Q_{\; \gamma,l}^{k}\Bigr)\biggr]Q_{\; \beta}^{l}
\; {\widehat G}^{\alpha \beta} \nonumber \\
&+& \Bigr(Q_{\; \alpha}^{j} \; F_{\; l}^{p}
-Q_{\; \alpha}^{p} \; F_{\; l}^{j}\Bigr)Q_{\; \beta,r}^{l} \; 
Q_{\; \gamma}^{r} \; {\widehat G}^{\alpha \beta} \nonumber \\
&+& \Bigr(Q_{\; \alpha}^{j} \; F_{\; l}^{p}
-Q_{\; \alpha}^{p} \; F_{\; l}^{j}\Bigr)Q_{\; \beta}^{l}
\Bigr(C_{\; \gamma \delta}^{\alpha} \; {\widehat G}^{\delta \beta}
-{\widehat G}_{\; \delta}^{\alpha} \; C_{\; \gamma}^{\delta \; \; \beta}
\Bigr)\biggr \} \delta \xi^{\gamma}.
\label{(5.4)}
\end{eqnarray}
In the particular (but relevant) case of Maxwell theory, $U^{jp}$ is
therefore gauge-invariant because $Q_{\; \alpha,j}^{i}=0$ as we said after
(4.6), and the structure constants $C_{\; \beta \gamma}^{\alpha}$ vanish
in the Abelian case.

Note also that the definition (2.1) can be generalized according to
\begin{equation}
[A,B] \equiv i{\hbar}(A,B)+\mu_{1}i{\hbar}A_{,j}U^{jk}B_{,k}
+\mu_{3}(i{\hbar})^{3}A_{,jkl}U^{jp}U^{kq}U^{lr}B_{,pqr}
+{\rm O}({\hbar}^{5}),
\label{(5.5)}
\end{equation}
where, for $A$ and $B$ obeying (1.7), $A_{,j}U^{jk}B_{,k}$ vanishes
for all type-I theories, by virtue of (4.1), whereas higher-order
terms only vanish in the Abelian case. With our notation,
${\rm O}({\hbar}^{5})$ denotes a finite (or possibly infinite)
number of contributions, of odd degree $\geq 5$ in ${\hbar}$.

As far as antisymmetric bilinear maps are concerned, the definition (5.5)
might be further generalized along the lines suggested in 
Ref. \cite{Espo07}, i.e. by including
$$
A \; {\rm exp} \left[{i{\hbar}\over 2}{{\lvec \delta}\over \delta
\varphi^{j}} {\widetilde G}^{jk} {{\rvec \delta}\over \delta \varphi^{k}}
\right]B 
-B \; {\rm exp} \left[{i{\hbar}\over 2}{{\lvec \delta}\over \delta
\varphi^{j}} {\widetilde G}^{jk} {{\rvec \delta}\over \delta \varphi^{k}}
\right]A.
$$
The formal expansion of this map yields
$$
i{\hbar}(A,B)+{(i{\hbar}/2)^{3}\over 3!}V_{AB}
+{\rm O}({\hbar}^{5}),
$$
where, on defining
\begin{equation}
W_{l}^{\; k}(P) \equiv (P_{,j}{\widetilde G}^{jk})_{,l}, \; P=A,B, 
\label{(5.6)}
\end{equation}
one finds
\begin{equation}
V_{AB}=-2W_{l}^{\; k}(A)\Bigr({\widetilde G}_{\; \; \; ,n}^{lm} \;
W_{m}^{\; n}(B)\Bigr)_{,k}
-2W_{l}^{\; k}(A)_{,n}\Bigr({\widetilde G}^{lm}W_{m}^{\; n}(B)\Bigr)_{,k}.
\label{(5.7)}
\end{equation}
It is therefore clear that higher orders in $\hbar$ bring in infinitely
many functional derivatives of the supercommutator 
${\widetilde G}^{jk}$. This is certainly interesting in the investigation
of the most general antisymmetric bilinear map, but not obviously useful
if one wants to obtain eventually a Poisson bracket on 
gauge-invariant functionals (cf. the important work in Ref. \cite{Hirs02}).

\section{Concluding remarks and open problems}

The Peierls bracket \cite{Peie52, DeWi65, Bimo03} 
has been applied in the modern literature on the manifestly covariant 
\cite{DeWi60, DeWi84, Nels86, Crnk87, Barn91, Maro94, Kana01, 
Duts03, More04, Ozak05} approach to quantization of gauge theories, 
including gravity, and some authors have even gone so far as to suggest 
that the Peierls bracket can be used to actually {\it define} the 
functional integral itself \cite{More04}. 

On the other hand, in ordinary quantum mechanics, the Poisson bracket can
be obtained from the first-order (in $\hbar$) expansion of the Moyal
bracket, and hence we have tried to understand 
whether the Peierls map for gauge theories can be suitably
generalized. Contrary to our original expectations \cite{Espo07}, the
extension here proposed is not of the Moyal type, since we have not studied
the phase-space formulation of type-I gauge theories, but rather their
space-of-histories formulation.

Our findings are expressed by the definitions (2.1) and (4.1): the geometry
of the space of histories for type-I gauge theories makes it possible to
obtain an antisymmetric bilinear map that is richer than the Peierls
map. At that stage, restriction to gauge-invariant functionals of the fields
reduces (2.1) to the Peierls bracket only in the case of Maxwell theory.
Our construction is richer than Peierls's if one just looks at antisymmetric
bilinear maps, but is considerably weaker if one looks for Poisson
brackets on the space of gauge-invariant field functionals.
At least two outstanding problems are therefore in sight:
\vskip 0.3cm
\noindent
(i) How to improve the definition (2.1) so that it gives a Poisson bracket
different from the Peierls bracket for all type-I gauge theories, when
restricted to the space of gauge-invariant functionals. Should one instead
look at maps having the general form (cf. Ref. \cite{Grac02} in ordinary
quantum mechanics)
\begin{equation}
[A,B](\varphi) \equiv \int L(\varphi,\chi,\psi;{\widetilde G})
(A(\chi)B(\psi)-B(\chi)A(\psi))d\mu (\chi,\psi),
\label{(6.1)}
\end{equation}
where $d\mu(\chi,\psi)$ is a measure on the space of histories, and try
to work out the form of the kernel $L(\varphi,\chi,\psi;{\widetilde G})$? 
\vskip 0.3cm
\noindent
(ii) Suppose one starts instead from a phase-space formulation of type-I
gauge theories. Within this framework, the formal analogy with ordinary
quantum mechanics on phase space might be exploited to find a suitable
Moyal bracket, that should be 
proportional to the Peierls bracket to lowest
order in $\hbar$. For this purpose, we plan to study first some examples
borrowed from classical and quantum dynamics with just one pair of
$(q,p)$ variables \cite{Espo08}, to begin with. 
Can the resulting Moyal bracket be re-expressed in
terms of position variables only? Does this shed new light on our goal
of generalizing the Peierls bracket?  

There is therefore room left for a lot of further work, and the hope remains 
that the space-time approach to quantum field theory \cite{DeWi84} might be
extended so as to understand what is a deeper foundation of Peierls
brackets \cite{Peie52} and gauge-invariant commutators \cite{DeWi60}.

\acknowledgments

We are indebted to G. Marmo for inspiring conversations, and to
J. Gracia--Bondia and P. Vitale for correspondence.
Our work has been partially supported by PRIN {\it SINTESI}.


\begin{references}
\bibitem{DeWi03}
B.S. DeWitt {\it The global approach to quantum field theory},
{\it International Series of Monographs on Physics} {\bf 114}
Clarendon Press, Oxford, 2003 (IMPHA,114,1); 
B.S. DeWitt in: {\it 50 Years of
Yang--Mills theory}, ed. G. 't Hooft, World Scientific, Singapore, 2005.
\bibitem{Espo07}
G. Esposito and C. Stornaiolo, hep-th/0607114.
\bibitem{Moya49}
J.E. Moyal, {\it Proc. Camb. Phil. Soc.} {\bf 45} (1949) 99.
\bibitem{Jord61}
T.F. Jordan and E.C.G. Sudarshan, {\it Rev. Mod. Phys.} {\bf 33}
(1961) 515.
\bibitem{Grac02}
J.M. Gracia--Bondia, F. Lizzi, G. Marmo and P. Vitale, 
{\it JHEP} 0204 (2002) 026.
\bibitem{Peie52}
R.E. Peierls, {\it Proc. R. Soc. Lond.} {\bf A214} (1952) 143.
\bibitem{Fadd67}
L.D. Faddeev and V.N. Popov, {\it Phys. Lett.} {\bf B25} (1967) 29.
\bibitem{DeWi67}
B.S. DeWitt, {\it Phys. Rev.} {\bf 162} (1967) 1195.
\bibitem{Hirs02}
A.C. Hirshfeld and P. Henselder, {\it Am. J. Phys.} {\bf 70}, 537 (2002);
{\it Ann. Phys. (N.Y.)} {\bf 298}, 382 (2002).
\bibitem{DeWi65}
B.S. DeWitt, {\it Dynamical theory of groups and fields},
Gordon \& Breach, New York, 1965.
\bibitem{Bimo03}
G. Bimonte, G. Esposito, G. Marmo, C. Stornaiolo,
{\it Int. J. Mod. Phys.} {\bf A18} (2003) 2033. 
\bibitem{DeWi60}
B.S. DeWitt, {\it Phys. Rev. Lett.} {\bf 4} (1960) 317. 
\bibitem{DeWi84}
B.S. DeWitt, {\it The spacetime approach to quantum field theory},
in {\it Relativity, Groups and Topology II}, eds B.S. DeWitt
and R. Stora, 381-738, North--Holland, Amsterdam, 1984.
\bibitem{Nels86}
J.E. Nelson and T. Regge, {\it Ann. Phys. (N.Y.)} {\bf 166} (1986) 234.
\bibitem{Crnk87}
C. Crnkovic and E. Witten, in: {\it Three Hundred Years of Gravitation},
eds. S.W. Hawking and W. Israel, Cambridge University Press,
Cambridge, 1987.
\bibitem{Barn91}
G. Barnich, M. Henneaux and C. Schomblond, {\it Phys. Rev.}
{\bf D44} (1991) 939.
\bibitem{Maro94}
D. Marolf, {\it Ann. Phys. (N.Y.)} {\bf 236} (1994) 374; {\bf 236} (1994) 392. 
\bibitem{Kana01}
I.V. Kanatchikov, {\it Int. J. Theor. Phys.} {\bf 40} (2001) 1121.
\bibitem{Duts03}
M. D\"{u}tsch and K. Fredenhagen, {\it Commun. Math. Phys.} {\bf 243}
(2003) 275.
\bibitem{More04}
B.S. DeWitt and C. DeWitt-Morette, {\it Ann. Phys. (N.Y.)} 
{\bf 314} (2004) 448.
\bibitem{Ozak05}
H. Ozaki, {\it Ann. Phys. (N.Y.)} {\bf 319} (2005) 364.
\bibitem{Espo08}
G. Esposito, G. Marmo and C. Stornaiolo, work in preparation.
\end{references}
\end{document}